# Depressed Surface Magnetization in Thin $La_{0.8}MnO_{3-\delta}$ Films

M. A. DeLeon[1], T. A. Tyson[1,*], C. Dubourdieu[2], A. Bossak[2], J. Dvorak[3], G. Bonfait[4]

[1]Department of Physics, New Jersey Institute of Technology, Newark, NJ 07102
[2]Laboratoire des Matériaux et du Génie Physique, CNRS UMR 5628, ENSPG 38402 St.Martin d'Héres, France
[3]Department of Chemistry, Brookhaven National Laboratory, P.O. Box 5000, Upton, NY 11973-5000, USA
[4]Department of Chemistry, Nuclear and Technological Institute, P-2686 Sacavem, Portugal

**Abstract**

A systematic study the magnetization in $La_{0.8}MnO_3$ films of thickness varying from ultra-thin to bulk-like has been conducted. The study reveals suppression of the bulk magnetization for films up to at least 1600 Å. In addition, the surface (top ~50 Å) of all films exhibits depressed magnetization as observed by x-ray magnetic circular dichroism (XMCD). The reduced surface magnetization is attributed to the coexistence of magnetic and nonmagnetic components of the same chemical composition.

**PACS:** 75.70.Rf, 75.47.Lx, 78.70.Dm, 75.70.Ak

Corresponding Author: tyson@adm.njit.edu

## I. Introduction

The lanthanum manganites are of interest not only for their potential applications in information storage devices but also from the scientific perspective due to the intriguing coupling of electronic, magnetic and structural parameters. The metallic state in these materials has been described by the so-called double exchange mechanism as proposed by Zener and further developed by de-Gennes[1,2] There is much interest in how the coupling of the various degrees of freedom are influenced by strain induced by lattice-substrate mismatch. Millis *et al.* showed that the Curie temperature is extremely sensitive to biaxial strain and that the reduction is quadratic in the Jahn-Teller distortion.[3]

The modifications of magnetic and transport properties with changes in thickness are of interest. In early work on manganite films, Jin *et al.* studied thickness effects on magnetoresistance for La-Ca-Mn-O films deposited on $LaAlO_3$ (LAO).[4] They have found an optimal thickness for magneto-resistance (MR) at about 1000 Å with decreased MR for thinner and, interestingly, thicker films. This dependence of thickness reveals an optimal strain for the magnetic and transport properties. Kwon *et al.* found similar thickness dependence for $La_{0.7}Sr_{0.3}MnO_3$ (LSMO) films deposited on LAO. Magnetic force microscopy measurements show a thickness dependence of a maze pattern of magnetic orientation first appearing at about 480 Å which disappears by 2500 Å thickness.[5] Jin *et al.* argued that thick films may consist of a highly strained, high-resistivity, and high-MR portion near the substrate and a less-strained low-resistivity, low-MR portion far from the substrate Herranz *et al.* found evidence of a nonconductive region at the film/substrate interface regardless of lattice mismatch or film thickness of $La_{0.67}Ca_{0.33}MnO_3$ thin films.[6] Ziese *et al.* conducted studies on $La_{0.7}Ca_{0.3}MnO_3$ films that are found to be homogeneously strained. Films with thickness over 70 nm show characteristics similar to that of bulk, but thinner films are ferromagnetic and insulating for much of the temperature range. Orbital ordering in these ultra-thin films is suggested to be the origin of the insulating state.[7] Chen *et al*. have found percolative behavior in thin LSMO films as evidenced from resistive relaxation, suggesting phase separation.[8] Recent resonant inelastic x-ray measurements reveal that substrate induced strain in thin $La_{0.8}MnO_{3-\delta}$ films results in 3d band localization.[9]

In previous studies of $La_{0.8}MnO_{3-\delta}$ films, we found an increase in the degree of local distortion of the $MnO_6$ octahedra with reduced film thickness by near edge x-ray absorption spectroscopy at the manganese K-Edge.[10, 11] Bulk magnetization measurements conducted by SQUID magnetometry on 60, 300, and 1600 Å thick films showed a reduction of $T_c$ and magnetic saturation moment with a reduction in thickness. Our previous works quantifies the strain from the substrate-film interface which may have an impact on the magnetic, structural and transport properties (see Table 1 in Ref. 10). The strain in the thinner films relative to the 1600 Å film is ~1%. The reduced magnetization may originate from spin canting due to homogeneous strain modifying the Mn-O-Mn bond angles or multiple magnetic/structural phases as induced by interface strain.[12] Transport measurements revealed good correlation of resistivity peak temperatures to Curie temperatures providing evidence of coherent conductive paths at the surface of the 60 and 1600 Å films measured.

In order to understand the origin of the reduced magnetic moment, we conducted a study comparing the surface magnetic moment to that of the volume averaged magnetic moment (SQUID measurement) of the films. To probe the surface magnetization, x-ray magnetic circular dichroism measurements (XMCD) were performed at the Mn L-edges. Analysis of the transport measurements at low temperature reveals that the 1600 Å film behaves similar to bulk LCMO[13,14], while the 60Å has an anomalous upturn at low temperature. Comparison of the bulk magnetic moment per manganese for 60, 300 and 1600 Å films with that of the 5000 Å film reveals that the magnetization saturation values of these films are depressed (even in the 1600 Å film) and varies significantly with thickness. On the other hand, The Mn L-Edge XMCD measurements, which probe the surface to a depth of ~50 Å reveal that the moment per Mn site at the surface is qualitatively the same in all films (60, 300 and 1600 Å). The results indicate that the surface of the films is most likely composed of multiple components which are chemically similar but with different magnetic ordering (ferromagnetic and nonferromagnetic components).

## II. Experimental Methods

Epitaxial $La_{1-x}MnO_{3-\delta}$ (x ~ 0.2) films were grown on (012) $LaAlO_3$ substrates (referred to as $LaAlO_3$ (001) by the pseudocubic designation) by metal-organic chemical vapor deposition (see Ref.. 10). A liquid-injection delivery scheme was used with a single liquid source. The β-diketonates La-(tmhd)$_3$ and Mn-(tmhd)$_3$ were dissolved together in 1,2-dimethoxylethane solvent. Proper La:Mn precursor ratio for the goal stoichiometry were determined by a series of depositions of varying precursor ratios in search of optimal magnetic properties for thick films. The process by which La:Mn ratios in the deposited films are determined relies upon previous works and magnetic profiles for ultra-thick films.[15] Additional information on the reactor and the experimental procedure are described elsewhere in Ref. 15 . Deposition runs were carried out at 700 ºC under a total pressure of 0.67 kPa and an oxygen partial pressure of 0.33 kPa. After deposition, the films were *in situ* annealed in one atmosphere of oxygen at 800 ºC for 15 minutes. Films of thickness 5000, 1600, 300, and 60 Å were prepared. The thickness was *in situ* monitored by a laser reflectometry set-up. The growth rate was ~ 2 Å per injection pulse with an injection rate of 1 Hz. The crystal, surface, and local structure were analyzed in a prior paper.[10,11]

Magnetization measurements were conducted by SQUID magnetometers in a 0.2T field in plane of the films and $T_c$'s are 220, 250, and 260K for the 60, 300, and 1600 Å films, respectively. (The magnetization values here are a correction to those the previously reported in (Ref. 10).) Resistivity measurements were conducted using a 4-contact in-line probe. The $T_p$ (peak resistivity) values for the films were 225, 255, and 260K for the 60, 300, and 1600 Å films, respectively. The low temperature resistivity (4.2K-40 K) of the films was fitted by methods as explained below.

X-ray magnetic circular dichroism (XMCD) spectra measured at the Mn L-edges were conducted on UV beamline U4B at the National Synchrotron Light Source (NSLS) at the Brookhaven National Laboratory (BNL). The degree of circular polarization was 70%(±5%). The films planes were oriented 45º to the incident x-ray beam and the magnetic field was applied in the sample plane. All XMCD spectra were measured at 90 K in remnance (after pulsing to ~0.1T). Magnetization (M) vs. applied field (H) hysteresis measurements at 100 K show that this field is sufficient to achieve saturation. The M vs. H loops (SQUID) were square for the 1600 and 300 Å thick films, but due to the large diamagnetic moment of the substrate, measurements on the thinnest films were indeterminate. In total electron yield (TEY) mode, this element specific technique supplies a comparison of the magnetically active manganese with a sampling depth (1/e) within the top ~50 Å of these three films.[16,17,18] High resolution measurements were conducted in the 60, 300 and 1600 Å films and reported in Ref. 10. The strain in the thinner films relative to the 1600 Å film is ~2%. Mn $K_\beta$ x-ray emission measurements (see Fig. 1) on the 60 and 1600 Å films show that the valence of Mn is the same in both films. (Fig. 1 can be compared with Ref. 19 for trends in Mn $K_\beta$ x-ray emission with Mn valence in manganites.). No evidence of $Mn^{2+}$ (as sharp shoulders on the main peaks) from additional chemical species is found in the Mn L-edge or XMCD spectra.[20] The samples exists in a single chemical phase.

### III. Results and Discussion

We have analyzed resistivity data in order to evaluate the transport mechanism.[10,11] Resistivity data in Fig. 2(a) show resistivity peak temperatures very close to their respective Curie temperatures (Fig. 2(b)). Fits at lower temperatures (Fig, 3), below 40K, result in good matches for metallic behavior for the 1600Å thick film, but due to an upward turn at very low temperatures, the 60Å has a poorer quality fit.

Zhao *et al*. have shown that low temperature resistivity in metallic manganites may be fitted by a small polaron model (PM) given by $\rho = \rho_0 + E \omega_s / \sinh^2(\hbar \omega_s / k_b T)$.[21] In this equation $\rho_0$ is the impurity scattering term, E is a constant proportional to the effective mass of the polaron and $\hbar \omega_s$ is the average energy of the softest optical mode. We have fitted our low temperature resistivity to this equation in Fig. 3(a) and found a good fit for the 1600Å film but not for the 60Å film. These polaron model fits are indicated as PF60 and PF1600 for the 60 Å and 1600 Å films, respectively, in Fig. 3.

Fits to an electron scattering (ESM) model were also conducted. For the 1600 Å the best fit was obtained with this model in which the resistivity $\rho$ is approximated by $\rho_0 + AT^2 + BT^{9/2}$ (where the $T^2$ term corresponds to electron-electron scattering and the $T^{9/2}$ term corresponds to electron-magnon scattering). These scattering model fits are indicated as SF60 and SF1600 for the 60 Å and 1600 Å films, respectively. Again, fits to the 60 Å film data with this model also failed. The increase in the low temperature resistivity in the thin film may be due an increase in the volume of a non-conducting phase resulting in increased percolative transport.[22] Of note, the low temperature upturn for the 60 Å thin film resistivity is similar to what is observed in polycrystalline films.[23, 24, 25] Dubourdieu *et al.* have measured such an upturn in low temperature resistivity data on polycrystalline films of $La_{1-x}Sr_xMnO_3$ on Si (001). The upturn in resistivity is attributed to either low energy inter-grain barriers or localization of carriers at low temperature. Note that LAO substrates often exhibit twinning, which may result in a mosaic of orientations for deposited film. Evidence of these orientations have been previously reported in phi-scans from x-ray diffraction.[10, 11] Another possibility is that the re-entrant insulating character may also be due to an increasing effect of lanthanum deficiency point defects on the transport properties of the thinnest film. Lanthanum point defects would have a larger effect on films of lower dimensionality. The most likely origin of the upturn in resistivity may be pressure/strain induced stabilization of an insulating phase. Recent, work on bulk samples of $La_{0.8}MnO_3$ reveal that this materials becomes insulating for hydrostatic pressures above ~4 GPa.[26] In addition, resonant inelastic x-ray scattering measurements reveal evidence of 3d band localization in thin films of this system (see Ref. 29).

Magnetization measurements of the three films show a systematic decrease of $T_c$ with decreasing film thickness (Fig. 2(b)). The 5000 Å thick film attains the theoretical magnetic saturation moment (3.4 $\mu_B$/Mn site), whereas the thinner films have reduced magnetization as well as $T_c$. Even the 1600 Å film with a $T_c$ of about 270 K has a reduced magnetization. The reduced magnetization may be due to homogeneously strain-induced canting of spins throughout the films or from coexistence of multiple magnetic phases. With respect to a phase separation model, we must observe that the 1600 Å is composed of connected metallic regions as ascertained from the resistivity measurements (Fig 2(a)). Metallic behavior exists even in the 60 Å film for a large portion of the temperature region. In addition, in the thin film, island-like growth is found by atomic force microscopy (Fig 2(a) inset). The low temperature upturn is consistent with a second insulating phase present in the thin film. In order to further understand the magnetic trends with thickness, we examined the magnetization with a smaller probing depth.

X-ray magnetic circular dichroism has several advantages over SQUID magnetometry, including element specificity. Hence we conducted Mn L-Edge measurements on the film samples. In Fig. 4, one can see that all the XMCD spectra are similar in shape (showing no chemical variation) and that the 1600 Å and 300 Å films have equivalent magnitudes. The discrepancy in difference spectra (inset of Fig. 4) for the 60 Å film in comparison to that of the other two films is less than 8% (as calculated by area up to 650.1 eV, the cutoff for the L3 edge). The similarity of the surface magnetization bears a marked difference with the bulk magnetization as measured by SQUID magnetometry.

Based on the sampling depth of XMCD in electron yield mode, the magnetization of the 60Å film as measured by SQUID magnetometry should be close to the magnetization as measured by XMCD. Based on the similarity in XMCD amplitudes, we conclude that the magnetization at the top ~50 Å of the two thicker films is nearly equivalent to that of the 60Å film. We observer that the XMCD amplitudes are less than the amplitudes of those found in $La_{0.7}Sr_{0.3}MnO_3$ films (17-23%) and $La_{0.7}Ca_{0.33}MnO_3$ .[18, 27, 28], The magnetic moment per manganese for these systems should be similar to that of $La_{0.8}MnO_3$. Indeed, resistivity curve widths and saturation magnetization values in $La_{0.8}MnO_3$ are found consistent with those of $La_{1-x}Ca_xMnO_3$ and $La_{1-x}Sr_xMnO_3$ systems. Both the XMCD and the SQUID measurements probe same volume in the 60Å film. Based on the thin film as a calibration, we are also led to the conclusion that the surface magnetization of the 300 and 1600 Å films is depressed.

Our fluorescence Mn K-edge x-ray absorption spectra (which probes ~ $10^4$ Å) has the same shape for all samples except for a broad main line in the 60 Å film (Ref. 10) . The broad Mn K-edge of the thinnest film indicates local distortions could be due to a single homogeneous (canted) phase . The shape of the Mn L-edges and Mn L-edge XMCD are quite similar for these TEY measurements. In addition, no chemical shifts were found in the Mn K-Edge spectra (and no changes with thickness were found in the emission spectra, Fig. 1). Concerning the 60 Å thick film in a spin-canting model, the largely reduced magnetization would indicate very large canting angles resulting in a large effect on transport properties. The metallic behavior exhibited by the 60 Å thick film resistivity data, however, indicates that it is more likely that the film has nonmagnetic and magnetic components rather than canting of spins. Concerning the thicker films, a nonmagnetic phase (or phases) on the surface is most likely a phase with distorted $MnO_6$ octahedra (as observed in the Mn K-edge measurements) but with only small changes in the crystallographic structure.[29]

In summary, a systematic study of $La_{0.8}MnO_3$ films of thickness varying from ultra-thin to bulk-like has been conducted. The study reveals suppression of the magnetization below bulk values for films up to at least 1600 Å. X-ray magnetic circular dichroism measurements at the Mn L-edges reveal a depressed surface magnetization exists in thick films with relaxed strain. This top layer is composed of a metallic ferromagnetic component with transport characteristic of bulk samples as well as nonmagnetic component with the same chemical composition.

Data acquisition was performed at Brookhaven National Laboratory's National Synchrotron Light Source (NSLS), which is funded by the U. S. Department of Energy. This research is funded by NSF DMR-0512196, NSF INT-0233316, and CNRS/NSF project No. 14550.

**Fig. 1.** Mn $K_\beta$ x-ray emission measurements on the 60 and 1600 Å films show that the valence of Mn is the same in both films.

**Fig. 2.** Resistivity and magnetization as a function of temperature. (a) The resistivity of the 60 and 1600 Å are shown with AFM images as insets. Note that although island growth exists in the 60 Å film, metallic transport is observed. In (b) the bulk magnetization is observed to vary strongly with thickness. Only in the 5000Å films is the theoretical saturation moment achieved.

**Fig. 3.** Fits to the low temperature resistivity: (a) PF60 and PF1600 are fits on the 60 and 1600 Å thick film low-temperature resistivity data, respectively, according to the polaron model: $\rho = \rho_0 + E\omega_s/\sinh^2(C/2T)$ where $C=\hbar\omega_s/k_b$ and (b) SF60 and SF1600 are fits on the 60 and 1600 Å thick film low-temperature resistivity data, respectively, according to the scattering model: $\rho = \rho_0 + A T^2 + B T^{9/2}$.

**Fig. 4.** The Mn L-Edge XAS and XMCD at 90 K as a function of thickness with an expansion of the XMCD near 643 eV. Note that the XMCD for 300 and 1600 Å films are identical while the signal for the 60Å is only 8 % smaller. The XMCD (MCD) signal in % is given as the lower curve and the inset.

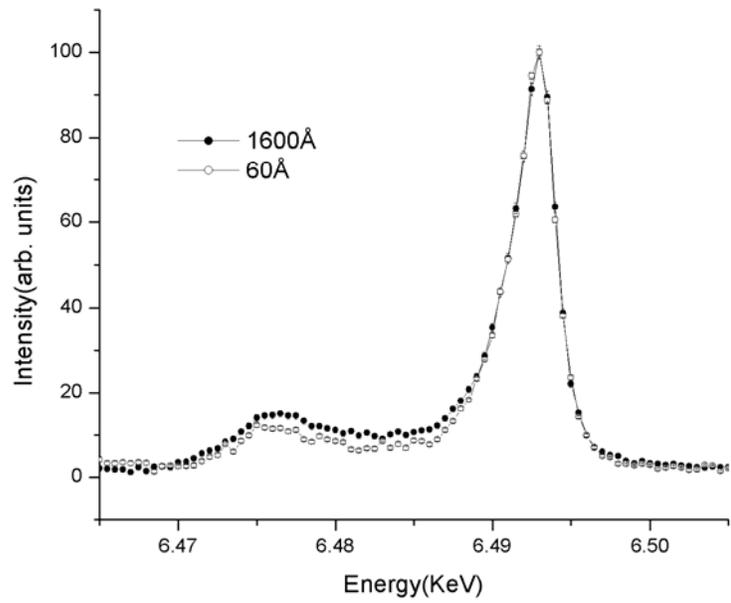
]
Fig. 1. (DeLeon *et al*.)

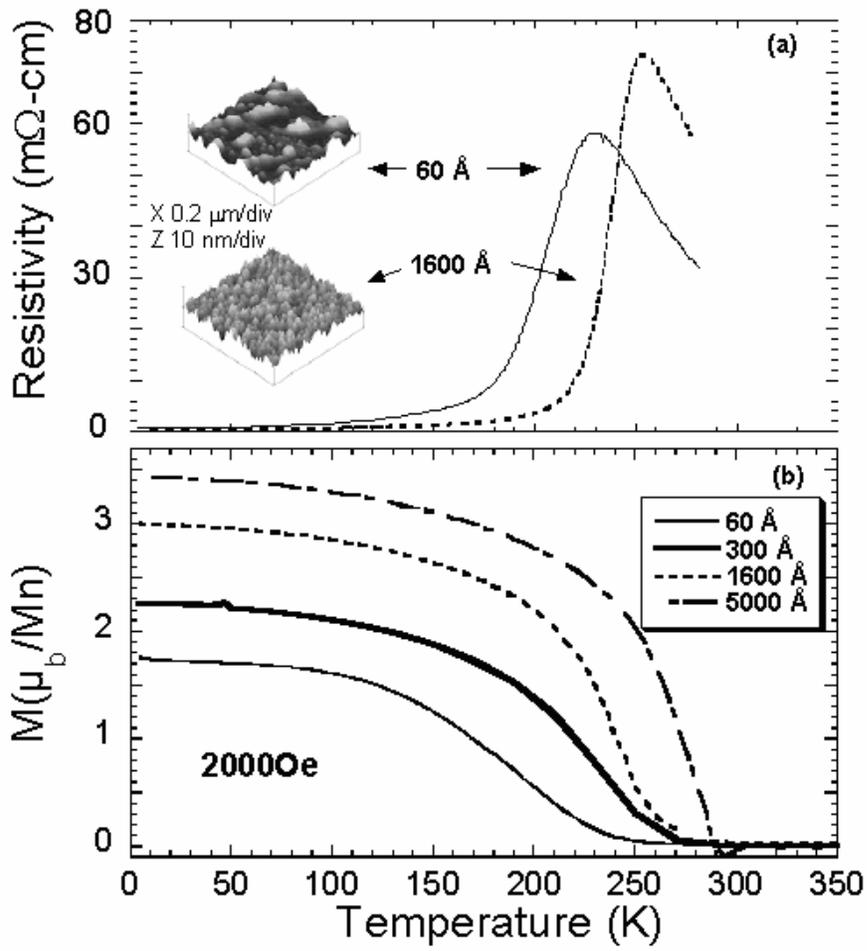

Fig. 2. (DeLeon *et al.*)

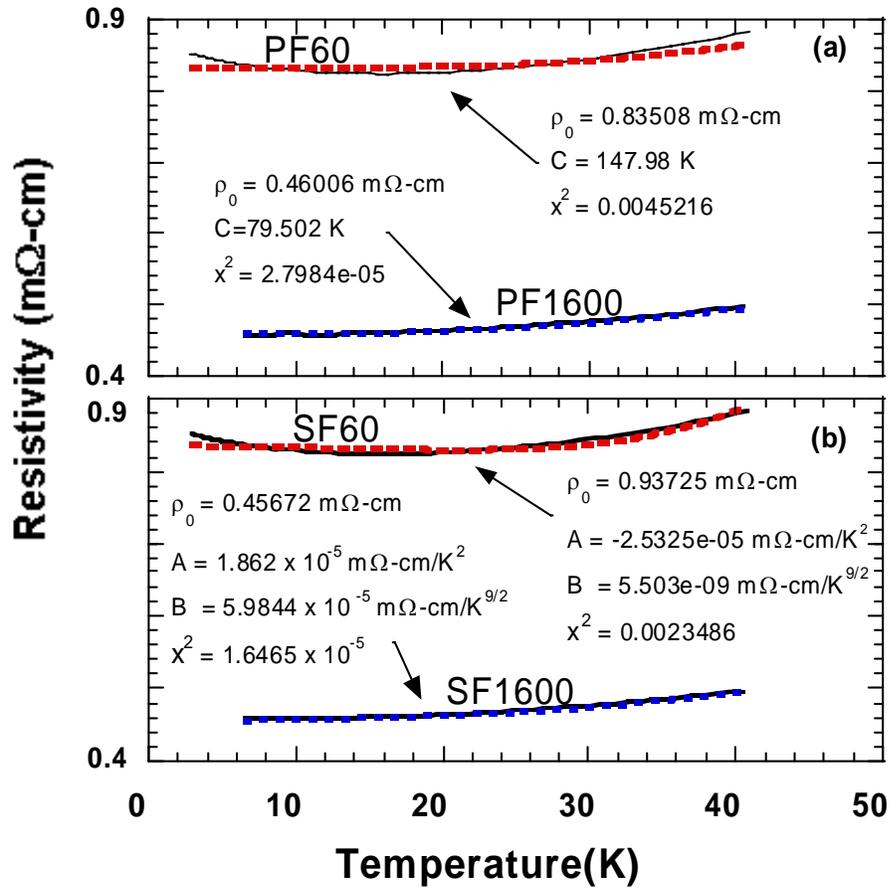

Fig. 3. (DeLeon *et al.*)

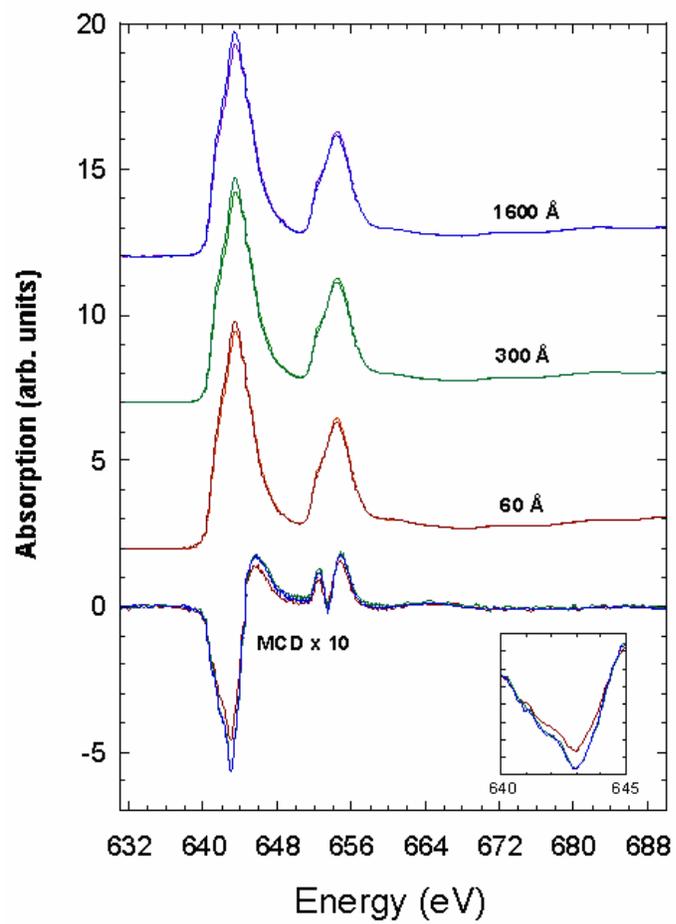

Fig. 4. (DeLeon *et al.*)